\documentclass[conference]{IEEEtran}
\IEEEoverridecommandlockouts
\usepackage{cite}
\usepackage{amsmath,amssymb,amsfonts}
\usepackage{algorithmic}
\usepackage{graphicx}
\usepackage{textcomp}
\usepackage{xcolor}
\usepackage{makecell}
\usepackage{url}
\usepackage{verbatim}
\makeatletter
\def\BibTeX{{\rm B\kern-.05em{\sc i\kern-.025em b}\kern-.08em
    T\kern-.1667em\lower.7ex\hbox{E}\kern-.125emX}}

\begin{document}

\title{Energy Efficient Scheduling of AI/ML Workloads on Multi Instance GPUs with Dynamic Repartitioning\\
}
\author{\IEEEauthorblockN{Ellie Lipe, Neel Karia, Connor Espenshade, Clifford Stein}
\IEEEauthorblockA{\textit{Columbia University}\\
New York, USA \\
\{eml2179, nmk2154, cje2136, cs2035\}@columbia.edu}
\and
\IEEEauthorblockN{Asser Tantawi, Olivier Tardieu}
\IEEEauthorblockA{\textit{IBM TJ Watson Research Center} \\
Yorktown Heights, USA \\
\{tantawi, tardieu\}@us.ibm.com}
}

\maketitle

\renewcommand\thefootnote{}
\footnotetext{\scriptsize Accepted version. Published in 2025 IEEE 25th International Symposium on Cluster, Cloud and Internet Computing (CCGrid), pp. 53--62. DOI: 10.1109/CCGRID64434.2025.00066. \\
© 2025 IEEE. Personal use of this material is permitted. Permission from IEEE must be obtained for all other uses, in any current or future media, including reprinting/republishing this material for advertising or promotional purposes, creating new collective works, for resale or redistribution to servers or lists, or reuse of any copyrighted component of this work in other works.}
\renewcommand\thefootnote{\arabic{footnote}}

\begin{abstract}
Increasing demand from AI/ML workloads is exacerbating the rising energy consumption of data centers. Recent advances in hardware such as NVIDIA’s Multi Instance GPUs (MIGs) offer improvements in flexibility and computational power and the opportunity for data centers to manage incoming jobs in energy-efficient ways, while maintaining acceptable performance. The challenge in achieving this multi-objective in a MIG environment through job scheduling is multi-faceted. Firstly, for a given MIG configuration, one seeks an easy-to-implement scheduling algorithm which selects a job from the queue as well as decides on which slice in the configuration the job runs. Secondly, for the identified scheduling algorithm, a particular MIG configuration may not always be suitable (as the workload fluctuates) and may need to be repartitioned. We tackle both problems using simulations and reinforcement learning (RL). We present a dynamic repartitioning scheduling framework for a single MIG as a solution to a multi-objective heterogeneous machine scheduling problem with preemption. In particular, we compare four scheduling algorithms and identify a promising one. Then, we employ reinforcement learning to perform dynamic repartitioning over a day. Furthermore, using a diurnal workload pattern based on real-world data center traces, we demonstrate the superiority of our dynamic repartitioning algorithm over twice-daily repartitioning ($26\%$), static partitioning ($31\%$) and no partitioning at all ($68\%$) according to a multi-objective function of energy consumption and tardiness. Our results indicate specific preferred configurations at different times of the day under different queue conditions, suggesting a policy for predictive and automatic reconfiguration.

\end{abstract}

\begin{IEEEkeywords}
Scheduling, Energy-Efficiency, Repartitioning, Tardiness, GPU, MIG, DQN, Reinforcement Learning, A100
\end{IEEEkeywords}

\newcommand{\cliff}[1]{\textcolor{violet}{\{Cliff: #1\}}}
\newcommand{\neel}[1]{\textcolor{brown}{\{Neel: #1\}}}
\newcommand{\asser}[1]{\textcolor{blue}{\{Asser: #1\}}}
\newcommand{\olivier}[1]{\textcolor{orange}{\{Olivier: #1\}}}
\newcommand{\ellie}[1]{\textcolor{magenta}{\{Ellie: #1\}}}
\newcommand{\connor}[1]{\textcolor{teal}{\{Connor: #1\}}}

\newcommand{\metric}[1]{\ensuremath{ET}}

\section{Introduction}
Data centers play a critical role in sustaining our modern world and enabling technological advances, but they currently carry a significant environmental cost. In 2022, data centers consumed $2\%$ of global energy demand with a total of 460 terawatt hours consumed\cite{datacenters}. This value is expected to more than double by 2026, an increase driven largely by the recent popularity of large language models (LLMs) and other AI/ML workloads, which depend on the use of 
power-hungry GPUs.

Efforts to reduce data center energy consumption include the development of advanced hardware and energy-efficient processors, widespread adoption of virtualization, the use of smart cooling systems, and strategic data center design. In addition to these developments, there is an immediate opportunity to further reduce energy consumption by scheduling workloads in a more efficient manner that considers the power characteristics of the GPUs.
In this work, we focus on the energy-efficient scheduling of energy-expensive AI/ML workloads. 

Recently developed GPU features present particularly interesting opportunities for scheduling algorithms. Companies such as NVIDIA have developed GPU-sharing features that help improve utilization by enabling the colocation and concurrent processing of multiple workloads on the same GPU. A recently released capability is the MIG feature. MIG enables the partitioning of its memory and compute resources to run multiple jobs in various {\em slices}, to utilize increasingly powerful GPUs more efficiently and avoid costly under-utilization. While MIG allows sharing one GPU among workloads, little has been done to recommend how to divide the GPU especially given energy-related objectives.  

There is often a trade-off between latency and energy usage when scheduling workloads \cite{ali2023performance}. The trade-off is well understood to be non-linear, as power consumption in computing devices is typically a super-linear function of their speed. 
We explore this trade-off for MIG-enabled GPUs, where workloads can be processed on different-sized slices of the GPU leading to different performance and energy consumption results. 

In this paper, we design an online scheduling framework that decides how to dynamically repartition a MIG-enabled GPU and allocate AI/ML workloads to its slices. We aim to achieve low energy consumption and good scheduling performance simultaneously, using a multi-parameter objective. 

The first novelty of our paper is its focus on energy-efficient scheduling for MIG, a relatively unexplored area that distinguishes our work from previous research on MIG scheduling problems. Previous work on MIG focused primarily on performance optimization, while our work explores scheduling with the dual objectives of energy and a quality-of-service metric. While our ultimate goal is to be able to schedule multi-GPU systems, an important step is to understand the single-GPU case. Thus, our work intentionally focuses on the single-GPU use case to establish a foundational understanding of energy-aware scheduling on MIG. 

The second novelty of our paper lies in the diverse AI/ML workloads we consider and the inelasticity of their scaling. Our queue consists of a mix of inference and fine-tuning or training jobs with varying elasticity; the performance of some jobs scales linearly with increased resources, while others scale sublinearly or not at all, adding significant complexity to our scheduling problem. We analyze an NVIDIA A100 40GB GPU as our single machine with an experimentally derived non-linear power curve across utilization levels. Although we focus on this setting, our methodology can extend, with appropriate modeling, to other computational environments and job mixes. This combination of a heterogeneous job mix, nonlinear elasticity, and nonlinear power along with preemption creates a complex problem that we solve for an online and realistic setting. In subsequent sections, we explain our modeling, algorithmic, and experimental choices.


Our MIG scheduling problem has similarities and key differences with other scheduling problems. The job scheduling problem within a MIG partition is related to scheduling with heterogeneous servers\cite{Kim2011}, where a slice corresponds to a server. In such a class of queueing problems, multiple types of jobs are served by a set of servers with different service rates. Threshold control policies have been designed (given some simplifying assumptions) and analytical results are limited to tractable cases\cite{Efrosinin2020}. However, the MIG scheduling problem is more complex in many ways discussed above, after including the diverse job mix and inelasticity. The re-partitioning problem is analogous to having a total service capacity that needs to be split among several servers with different capacities such that the sum of the capacities is equal to the total available capacity. To complicate matters, only a particular set of combinations are possible. 

This problem setup gives rise to three subproblems: (i) choosing the partitioning of the GPU from a fixed set of possible slicing profiles, (ii) deciding the prioritization of jobs from the queue, and (iii) selecting an available slice to assign to each job. The exact version of our scheduling problem is NP-hard, as it is NP-hard even without considering the energy and with a fixed configuration of MIG with more than one slice (which is implied by a result in \cite{sitters2005complexity}). We solve this problem in two parts: (1) scheduling jobs onto slices within a given partition and (2) deciding when to repartition and to which configuration to repartition. We address (1) in Subsection \ref{sec:within} using preemption-based scheduling algorithms, and (2) in Subsection \ref{sec:repartition} using a reinforcement learning-based approach (dynamic repartitioning). We summarize our results below:
\begin{itemize}
    \item We define a multi-objective metric named \metric~ in Subsection \ref{metric} that considers both energy consumption and average tardiness to relatively evaluate our algorithms.
    
    \item We test several algorithms and experimentally observe that the Earliest Deadline First - Slowest Slice (EDF-SS) algorithm generally outperforms the other algorithms for scheduling different queues of inelastic jobs within various MIG configurations.


    \item By using dynamic repartitioning with a reinforcement learning model based on the EDF-SS scheduling algorithm, we obtain considerable improvements according to \metric~. When evaluated based on a hybrid of real and simulated data, the RL-based dynamic repartitioning algorithm performs better than the benchmark models of twice-daily repartitioning by $26\%$, static partitioning by $31\%$, and no partitioning by $68\%$.
    
\end{itemize}

\section{Related Work}

There has been a plethora of research involving the scheduling of jobs on multiple machines or multiple cores (and more recently on multiple slices of a GPU) while trying to optimize a combination of a few of the following - cost, energy, latency, throughput, and accuracy. Recent work related to MIG has focused almost exclusively on online scheduling and performance-related objectives often for large clusters of GPUs \cite{paris-elsa,gpu_serving}. Clover \cite{li2023clover} considers another sustainability objective of carbon reduction while scheduling inference jobs with MIG by adjusting job parameters but does not incorporate GPU repartitioning. Our work differs from previous MIG scheduling work in its objective to reduce the energy consumption of diverse workloads by repartitioning the GPU. 




Energy-aware scheduling has received significant attention recently and some works have also considered energy-aware scheduling for AI/ML workloads, specifically the two closely related works from McDonald et al. \cite{mcdonald2022great}  and Zeus \cite{you2023zeus}. The former focuses on reducing the energy consumption of NLP workloads by setting \emph{power caps}, a way of restricting the maximum power usage of a GPU, and the latter includes both energy and throughput in their objective and introduces the Zeus optimization framework that dynamically adjusts the batch sizes and power caps on the GPU for DNN jobs. Our research, on the other hand, considers a variety of AI/ML workloads with varying throughputs and does not modify the power cap of the GPU or the job characteristics.


Reinforcement learning (RL) can be suitable for problems that require sequential decision-making and have complex and dynamic environments, and recent research has explored using different RL algorithms to effectively solve a variety of scheduling problems by selecting jobs and or their resources to optimize for various objective functions \cite{kayhan2023,mao2016resource}. RL has been applied once before to MIG scheduling problems, where \cite{Saroliya2023} used RL to combine MIG and MPS modes with the performance objective of improving throughput, but this work did not consider energy. Furthermore, RL has been explored in other energy scheduling problems; \cite{Song2021ADR} and \cite{yi2019}  optimized for data center energy consumption by selecting jobs from the queue to allocate to specific servers. The former used an actor-critic algorithm and scheduled jobs in a round-robin approach while the latter used Deep RL while considering the power and thermal characteristics of the servers. Our energy scheduling problem differs from previous scheduling work with RL, as we have an additional decision with MIG to select the appropriate GPU configurations. 

\section{Background}
This section introduces Multi-Instance GPUs (\ref{sec:background-gpus}) and AI/ML workload of interest~(\ref{sec:background-workloads}).  We describe the previous work on MIG in Subsection~\ref{sec:background-gpus}.  In Subsection~\ref{sec:power-char}, we describe some of our experiments on the power characteristics of MIG on an A100 and then use those insights in the remainder of the paper.

\subsection{Multi-Instance GPUs}
\label{sec:background-gpus}
Recent NVIDIA data center GPUs can support Multi-Instance GPUs (MIG). When in MIG mode, the system administrator can partition a GPU into up to seven discrete, isolated slices. Unlike NVIDIA's Multi-Process Service (MPS), MIG partitions all elements of a GPU instead of having groups of streaming multiprocessor (SM) cores sharing all other GPU resources, including one shared memory pool. As such, each MIG slice comes complete with its own SM cores, memory, cache, and memory bandwidth. These slices can run workloads concurrently without interference or concerns about improper data access. Enabling MIG mode and creating and destroying MIG slices is completely done through the APIs of the GPU driver. In contrast to MPS, MIG does not require CUDA library changes or an additional software service to orchestrate CUDA kernel calls across multiple workloads. 

\begin{figure}
\centering
  \centering
  \includegraphics[width=0.9\linewidth]{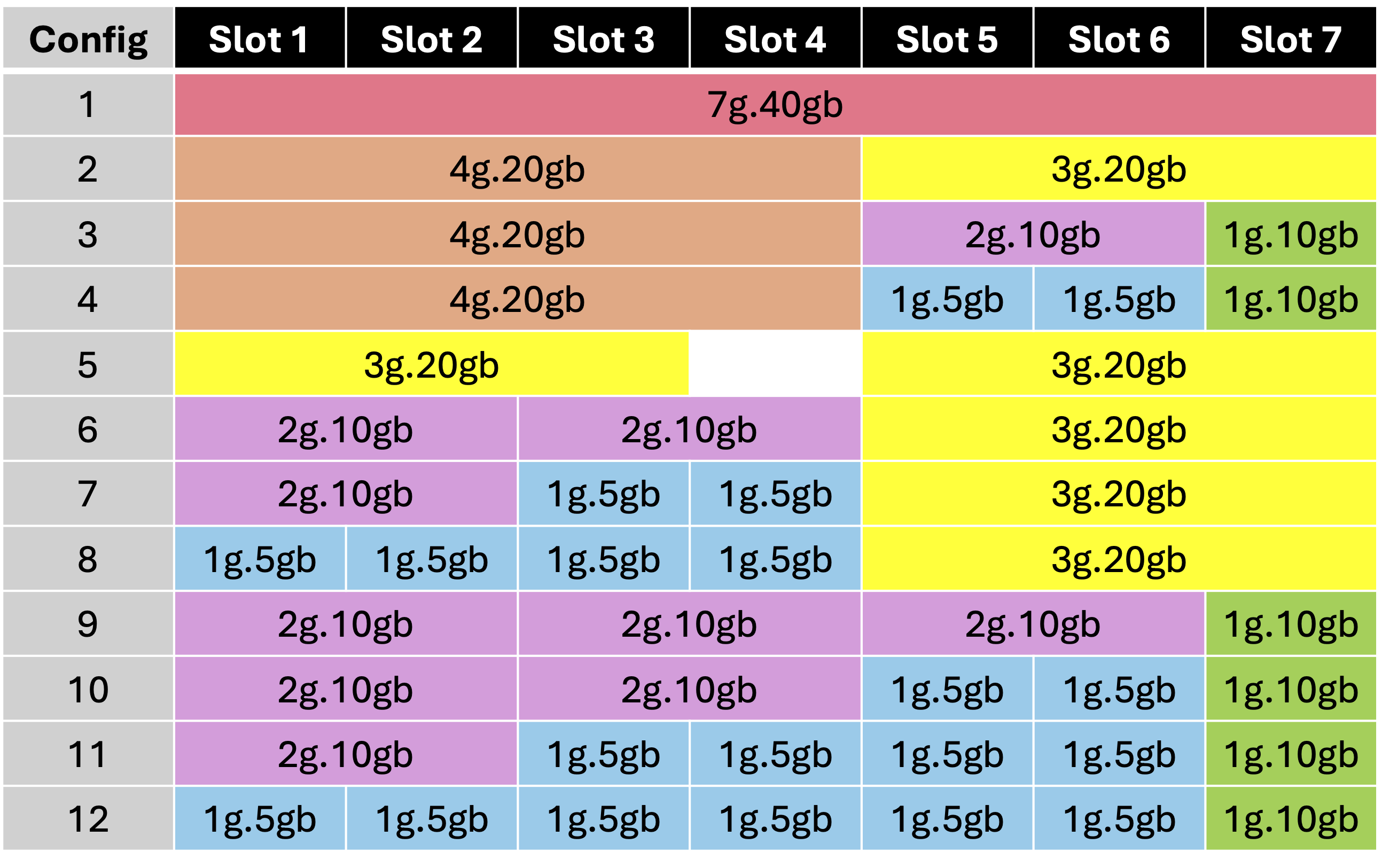}
  \caption{Slice Configurations of A100 40GB with MIG}
  \label{fig:slice_config}
\end{figure}

A MIG-enabled GPU supports several partitioning configurations, obtained by fusing none, some, or all of the seven possible MIG slices. The GPU only fuses slices in a manner which results in slice sizes of 1, 2, 3, 4, and 7. Figure~\ref{fig:slice_config} enumerates the configurations we consider on an A100-40GB GPU in this work. Although additional permutations of slices are possible, we focus on slices that leverage the full 40GB memory capacity available on the A100-40GB GPU and ignore other configurations featuring identical slices in differing orders. A two-slot-wide slice also known as a 2g slice has twice the amount of SM cores than a 1g slice, and so on.

Slices associate SM cores with memory. An A100 GPU with 40GB of memory supports the following types of slices: 1g.5gb, 1g.10gb, 2g.10gb, 3g.20gb, 4g.20gb, and 7g.40gb. These types reflect the supported combinations of SM core counts and memory sizes, e.g., 10 GB in a 2g.10gb slice. As a combination of two 3g.20gb slices consume the full amount of GPU memory, and configuration 5 has a hole, which cannot be occupied by another slice. In practice, the combination of a 4g.20gb and a 3g.20gb slices should be preferred. For similar reasons, at most one 1g.10gb slice should be included in any configuration. We do not consider configurations with multiple 1g.10gb slices, which underutilize SM cores.

Experiments in \cite{espenshade2024} find that running identical jobs on these isolated slices usually enables faster performance and lower energy per training sample processed compared to running on the entire GPU. 
Smaller convolutional neural networks with no more than $25$ million parameters such as DenseNet and ResNet experience an increase of training time per training sample of 10\% across all jobs compared to running these identical workloads sequentially. Further, power capping 
the GPU to $200W$ can result in energy $25\%$ lower than a non-MIG GPU, while still delivering faster performance. With some workloads such as fine-tuning, MIG can deliver up to $55\%$ faster performance across all jobs at $42\%$ less energy. As such, existing research purports that MIG can effectively increase performance and lower energy.

\subsection{Training and Inference Workloads}
\label{sec:background-workloads}

We analyze our scheduling algorithms on a mixed queue of inference and training workloads with varying throughput elasticities that can fit on an A100 40GB GPU. Note that we include Large Language Model (LLM) fine-tuning jobs in the training job category. Many past works restrict themselves to working with either training jobs or inference jobs.  While this restriction can simplify the problem, it ignores the potential efficiencies gained by scheduling the training and inference jobs together; because they have very different processing and memory usages, inference and training jobs can at times share a processor/slice that otherwise would not have the opportunity to handle multiple jobs at the same time.  
Even within the class of training jobs or of inference jobs, there are several types of jobs with very different performance characteristics; although such information is not known when a job first arrives at a data center, there are online throughput profilers, such as MISO \cite{Li2022}, that can measure and derive elasticities should this solution be implemented. In our work, we assume this information is available from a profiler, and attempt to model a realistic mix by including various elasticities.
\cite{li2022characterizing} and \cite{tan2021serving}  observe  
that the variation in the throughput with the size of the slice being used depends on the type of the job and the batch size, and is independent of the jobs in the other slices. (Table \ref{metric_eg}). 



\begin{figure}
\centering

\includegraphics[width=0.85\linewidth]{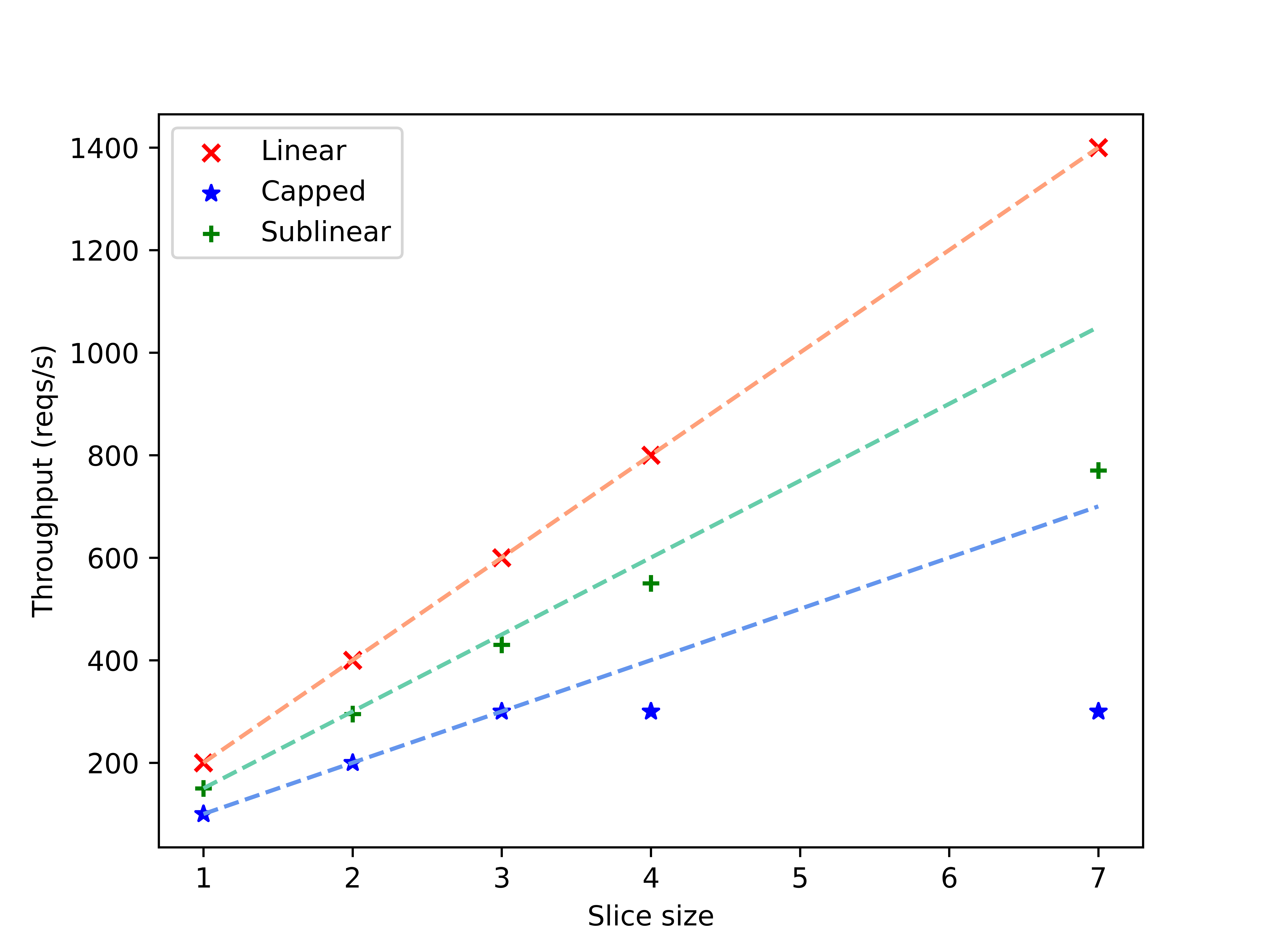}
\caption{Linear, Capped and Sublinear Throughputs}
\label{throughpur_char}
\end{figure}

We introduce a job classification scheme where the training and inference jobs can be placed into three categories based on their throughput scaling characteristics across MIG slices: linear throughput jobs, capped throughput jobs, and sublinear throughput jobs.  Refer to Figure \ref{throughpur_char} for an illustration of each of them.
\begin{itemize}
    \item \textbf{Linear Throughput jobs}: For these jobs, the throughput increases linearly with the increase in slice size.  Linear throughput jobs can be easily parallelized and take full advantage of additional computational power. 
    \item \textbf{Capped Throughput jobs}: For these jobs, the throughput first increases linearly with the increase in slice size and then remains constant.   Capped throughput jobs have a limited (capped) amount of parallelism and can benefit from some additional resources, but after some threshold cannot take further advantage. 
    \item \textbf{Sublinear Throughput jobs}: For these jobs, the throughput increases as a sublinear function of the slice size.   Sublinear throughput jobs can take advantage of parallelism but with a certain loss of efficiency.  It is still beneficial to give such jobs more slices, but they do not use the additional resources with the same efficiency as the linear throughput jobs. 
\end{itemize}
We have classified common AI/ML inference (I) and training (T) workloads into one of three categories based on the measurements from \cite{tan2021serving} and \cite{espenshade2024}. The throughput characteristics of these common AI/ML workloads are then incorporated into our experiments.


\begin{table}[ht]
  \caption{Throughput Categories for Common AI/ML Workloads }
  \label{metric_eg}
  \begin{center}
    \begin{tabular}
    {p{0.75cm}|p{1.91cm}|p{1.91cm}|p{1.91cm}}
        \hline
        Batch Sizes & Linear & Sublinear & Capped \\
        \hline
        1, \mbox{8} & Roberta-L(I) \mbox{Alberta-L(I)} \mbox{BERT(T)} & Embedding(T) \mbox{GPT2 Fine-tune} \mbox{Densenet169(T)} & Resnet(I) \mbox{Inception(I)} \mbox{Deepspeech(I)} \mbox{GNN(I)} \\
        \hline
        16, \mbox{32} & Resnet(I) \mbox{VGG19(I)} &  Densenet121(I) \mbox{InceptionV3(I)} & N/A \\
        \hline
    \end{tabular}
  \end{center}
\end{table}


\section{Our Proposed Solution}

To model the AI/ML workloads in a GPU, we consider a set of jobs, with some amount of processing. We use $\rm{dur}_{js}$ 
to denote the amount of processing a job $j$ would need on slice $s$.  As discussed in the previous section, the way the amount of processing depends on the slice varies according to the type of job and the particular slice.   We assume that a job $j$ has a deadline $d_j$, which either represents a user-specified (soft) deadline or a best-effort deadline assigned by the system.   A schedule will assign a job to certain slices at certain times and will lead to a job $j$ having a completion time $c_j$ and using a certain amount of energy.  The energy will be measured in the real system and can be modeled in our simulation.   Our quality of service measure is {\em tardiness}, and we define the tardiness of job $j$ with deadline $d_j$ and completion time $c_j$ as $l_j = \max(c_j-d_j, 0)$.  Tardiness explicitly measures the amount by which a job violates its deadline.   
 If a simulation has $m$ jobs, then the average tardiness of the simulation is $\sum_j l_j/m$. The solution is composed of two parts: first it employs a scheduling algorithm to allocate incoming jobs to the different available slices within a MIG configuration, and then it uses a reinforcement learning model to dynamically select which configuration it should use given the state of the job queue at that particular time. 
 
 The remainder of this section describes how these decisions are made. Subsection~\ref{metric}
describes the metric we choose. Subsection~\ref{sec:power-char}  describes the power characteristics of MIG on an A100 40GB GPU. The details we find do not follow the models in most of the literature (such as the ``power law" model which says that power is speed raised to a power $\alpha$, where $\alpha$ is a small constant like $2$ or $3$), but represent a more realistic understanding in terms of MIG. In Subsection~\ref{sec:within}, we discuss several algorithms that we use for scheduling jobs once we have chosen a MIG configuration. We experiment with multiple algorithms and then pick the one that performs the best empirically in most of the cases, recognizing that the decision could be different for different GPUs or different workloads.  Finally in Subsection~\ref{sec:repartition}, we discuss how and when to repartition the GPU.  Here, the algorithmic choices are quite complex, and hence we model them using a reinforcement learning approach.  

\subsection{Metric to be Used}\label{metric}

As we have two competing metrics, energy and tardiness, it is convenient to combine them into one measure.  There are several ways to do so; here 
 we define a weighted energy and tardiness multi-objective metric \metric~, which is similar to the metric used in \cite{10.1016/j.cie.2012.10.002}, and variants of it have often been used. Conceptually, it takes the weighted average of the energy and average tardiness for each simulation and then averages this over the $n$ simulations. It is given by $$\metric ~= \frac{1}{n}\sum_{i=1}^n\frac{ax_i + y_i}{a+1}$$ where $x_i$ is the energy and $y_i$ is the average tardiness for simulation $i$. $a$ is a scaling factor used to denote the relative importance of the tardiness to the energy. The lower the value of \metric~, the better the algorithm.

We choose $a$ as follows for consistency across all experiments. Let $s$ be the global mean of the energy across the simulations of all the experiments over the given algorithms and let $t$ be the global mean of the average tardiness across the simulations of all the experiments over the given algorithms. Then we set $a$ to be $\frac{t}{2s}$ 
. Note that the factor of $2$ yields a higher relative penalty for tardiness as compared to energy after the desired normalization.  

We use \metric~ as a standard to compare the performance of various algorithms described in the next subsection. The algorithm which performs the best is then used for the RL model.


\subsection{Power Characteristics of MIG}\label{sec:power-char}
\begin{figure}
\centering
\includegraphics[width=0.75\linewidth]{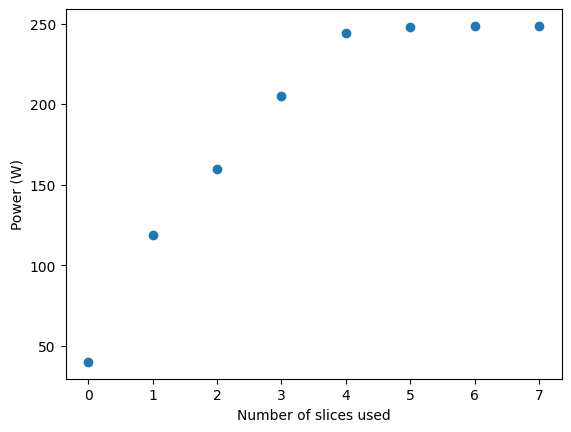}
\caption{Power Characteristics of A100 at 250W Cap}
\label{power_char}
\end{figure}

The interesting power characteristics of MIG that we observe help motivate our solution. We see that at the default $250W$ power cap, the power varies with the number of slices used as shown in Figure \ref{power_char}. Note that the variation in power between using multiple smaller slices summing up to a larger slice (for example three 1g slices vs one 3g slice) is never larger than $10\%$ (and generally less than $5\%$), so we ignore this for the ease of modeling. 
The power curve gives us the idea that we must try to utilize the GPU as much as possible for energy efficiency.

The power consumption per slice is high for the smaller slices and after using $4$ out of $7$ slices, we can run jobs on the rest of the GPU with a negligible increase in total power. This implies that for linear throughput jobs, the best solution would be to use the entire GPU. Hence, we should pack the jobs in the best possible way with highly increased throughput for a minimal increase in power. However, if there are sublinear and/or capped jobs present, then it turns out to be more nuanced (refer to Section \ref{sec:exp}).

\subsection{Scheduling Within a Configuration}
\label{sec:within}
Within a given MIG configuration, our solution must decide which jobs to prioritize from the queue and which slices to allocate them to. Given the power curve for MIG slices as well as the combination of sublinear, capped, and linear job performance 
profiles, we expect the algorithm to choose to pack the GPU with as many concurrent jobs and therefore maximize periods of overlap between active jobs while completing them on time. Packing the GPU will exploit the performance gains of multiple slices without significantly greater energy consumption.

We evaluate some common heuristic techniques for this scheduling problem. We use variants of the techniques that are better tailored to MIG and the heterogeneous machine problem setting. We consider this problem to be a heterogeneous machine scheduling problem because instead of scheduling jobs on one whole GPU, we are scheduling jobs on different slices of the GPU.


The algorithms evaluated follow a job prioritization logic of either Earliest Deadline First (EDF) or Least Laxity First (LLF). We define the laxity $l_{js}$ as the difference between a job $j$’s deadline $d_j$ and its remaining duration $\rm{dur}_{js}$ on slice $s$ so $l_{js}$ = $d_j$ - $\rm{dur}_{js}$. Because a job will have a different duration dependent on the slice it runs on, laxity must be consistently measured against a specified slice. This decision will vary between the least laxity algorithms and is explained in detail for both of them below. 

All algorithms allow for preemption, which is an important characteristic for online scheduling especially for AI/ML inference workloads with typically very short duration and tight deadlines. Also, we assume that the larger training jobs are sufficiently checkpointed and the cost of preemption is not very large. Preemption in an algorithm means that at specific points in time, all jobs currently in the system (either actively running at that moment or waiting in the queue) are re-evaluated for job prioritization and slice allocation. Any job can be preempted. The possible points of time for preemption are defined in each algorithm.

For slice allocation decisions, we may refer to the fastest or slowest slice for a particular job on a certain configuration as the slice that completes the job in the least or greatest amount of time, respectively. 

The scheduling algorithms neither guarantee that all jobs will be completed on time nor discard a job that might be impossible to complete on time. For example, if a job can only be completed on time if it runs on a 7g.40GB slice, but we are using a configuration that doesn't have this slice, the job will still run. How these jobs are handled depends on the algorithm and it is explained in detail below.



The following preemptive algorithms are evaluated: 

\begin{enumerate}
    \item \textbf{Earliest Deadline First – Fastest Slice (EDF – FS)}\\ Jobs from the queue are prioritized according to their deadlines and are always allocated to the fastest slice for the job. Preemption can occur when a job either arrives or completes and any job may be preempted.
    \item \textbf{Earliest Deadline First - Slowest Slice by Deadline (EDF – SS)}\\ Jobs from the queue are prioritized according to their deadlines and are allocated to the slowest slice which can complete it on time. Preemption can occur when a job either arrives or completes and any job may be preempted. If it is not possible for a job to meet its deadline on any of the configuration's slices, it will run on the fastest slice. 
    \item \textbf{Least Laxity First – Fastest Slice Laxity with Critical Laxity Events (LLF)}\\ Jobs are prioritized according to their laxity, which is measured against the duration on the fastest slice in the configuration for every job. Preemption can occur when a job either arrives, completes, or falls within a critical laxity threshold and within a maximum number of critical laxity preemptions per job. This maximum is set to help avoid throttling between two jobs with similar laxity values. Jobs are run on the fastest slice.
    \item \textbf{Least Average Laxity First – Fastest Slice with With Critical Laxity Events (LALF)}\\ Jobs are prioritized according to their average laxity, which is calculated by taking the mean duration across all slices in the configuration for every job. By consulting an average duration, jobs with different performance speedups between slices are better accounted for. Similar to LLF – FS above, preemption can occur when a job either arrives, completes, or falls within a critical laxity threshold and within a maximum number of critical laxity preemptions per job. This maximum is set to help avoid throttling between two jobs with similar laxity values. Jobs are run on the fastest slice.
\end{enumerate}

\begin{table}[ht]
  \begin{center}
    \begin{tabular}{p{3.5cm}|p{3.5cm}}
        \hline
        Algorithm & \metric~ \\
        \hline
        EDF-FS & 5.53 \\
        EDF-SS & 5.37 \\
        LLF & 6.32 \\
        LALF & 6.83 \\
        \hline
    \end{tabular}
  \end{center}
   \caption{\metric~ metric values for the four different algorithms. The value is averaged over multiple simulations.}
   \label{single_conf}
\end{table}

An algorithm can perform relatively better or worse depending on the different conditions, the different MIG configurations, and under different queue loads. We select EDF-SS as our scheduling algorithm to be considered within our repartitioning framework because it has the lowest performance metric overall as seen in Table \ref{single_conf}, and it most effectively packs the GPU relative to the other algorithms across most configurations. It also uses fewer preemptions than the least laxity algorithms. A deeper analysis of the algorithm's behavior will follow in Section \ref{sec:exp}.

\begin{figure}
\centering
  \includegraphics[width=\linewidth]{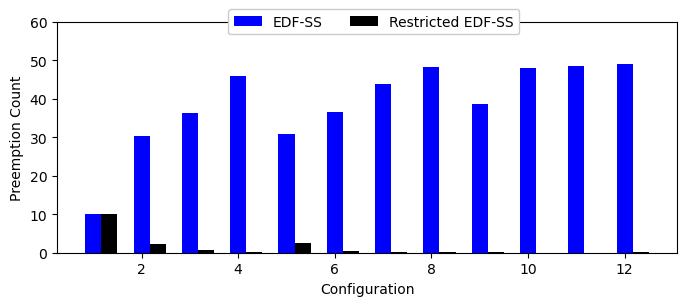}
  \caption{Preemption Frequency Across Configurations}
  \label{fig:pmtn_8005}
\end{figure}

While preemption is essential for online scheduling and we do not impose any penalty for it when scheduling within a single GPU, we recognize that unrestricted preemption can lead to unnecessary complexity, such as switching two jobs between different slices that might not result in any energy consumption or tardiness differences. Consequently, we investigate EDF-SS further and compare the original design that preempts without restriction against a variant that only preempts jobs in cases that directly prevent the jobs from missing their deadlines. The results in Figure~\ref{fig:pmtn_8005} show that EDF-SS with restricted preemption achieves similar \metric~ results to unrestricted EDF-SS but does so with a $63-99\%$ reduction in preemption occurrences. We therefore proceed with restricted EDF-SS as the algorithm that overall achieves the best \metric~ results with minimal preemptions, and further discussion on EDF-SS will focus on the restricted version.



\subsection{Re-Partitioning}
\label{sec:repartition}
Now that we can schedule within a given configuration, the second part of our solution addresses the challenge of selecting and re-selecting the configuration best suited to a particular queue. We employ a reinforcement learning model to determine when to repartition and which configuration to adopt. The MIG repartitioning problem is particularly well-suited for reinforcement learning due to its need for sequential decision-making and the inherent uncertainty of the environment given stochastic queue and job characteristics. This complexity necessitates an adaptive approach capable of learning from and reacting to dynamic variables. While we considered applying reinforcement learning for both repartitioning and job scheduling as a comprehensive solution, this would significantly expand both the state and action spaces, leading to the curse of dimensionality and complicating training convergence. Therefore, we opted for a two-part solution that utilizes Deep Q-Learning, incorporating an epsilon-greedy exploration-exploitation strategy and an experience replay buffer \cite{li2018deepreinforcementlearningoverview} with an event-driven architecture designed as follows:

\subsubsection{State Representation}
Our model's state representation concatenates $2+2m$ different features: the current MIG configuration, time, and the deadlines and average duration of the first $m$ jobs. The MIG configuration can take one of $12$ predefined values defined in \ref{fig:slice_config}. Since the other state features are naturally continuous, they are binned to discretize the state space. Note that increasing the value of $m$ will enlarge the state space and its complexity, directly affecting the model’s run-time and convergence. Thus, careful consideration is necessary when selecting an appropriate value. For this problem, we use $m = 3$ based on an analysis of typical GPU loads in Alibaba’s data center traces, discussed further in Section \ref{sec:exp}.

\subsubsection{Action Space}
The RL agent will choose to either remain on the current MIG configuration or to repartition to one of the $11$ other configurations. This constitutes an action space of $12$ different actions, corresponding to all the MIG configurations considered in this problem. Our model follows an event-based architecture where the agent is empowered to make decisions at certain points in time; re-partitioning can happen when a job arrives or is completed. 

\subsubsection{Reward Model}
Due to the multi-objective nature of our problem, our reward model is based on the \metric~ metric and the scalarization of energy and average tardiness with discounted rewards over time. Changing configurations incurs a performance penalty equivalent to the time required for the repartitioning process. This time is observed to be $4$ seconds (needed to destroy and recreate the necessary MIG slices for the chosen configuration).
 \\


\section{Experiments and Results}\label{sec:exp}


We conduct a series of experiments based on data derived from data center traces from Alibaba\cite{weng2022mlaas} to evaluate the results of EDF-SS with the other candidate scheduling algorithms, and present the results of dynamic repartitioning against benchmark models including a full GPU without MIG, MIG that remains fixed in one static configuration, and MIG that repartitions between day and night intervals. Our experiments and results show that EDF-SS is usually, but not always, the strongest scheduling algorithm across a variety of settings, and that dynamic repartitioning reduces \metric~ by 26\%-68\% compared to the benchmark models while improving both energy and performance objectives.  

\subsection{Simulation Overview} 
\begin{figure}[ht]
    \centering
    \includegraphics[width=\linewidth]{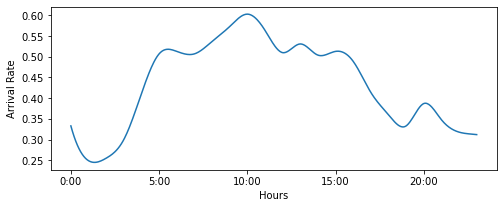}
    \caption{Variation of Arrival Rate Throughout 24-Hour Day}
    \label{arr_rates}
\end{figure}

We evaluate the in-configuration algorithms over 250 job queue iterations. We model the queues based on data from the Alibaba data center traces\cite{weng2022mlaas} and we make assumptions about information excluded from the traces that our problem requires as input. Job queues are simulated in the following way: jobs arrive according to a Poisson process with a nonhomogeneous arrival rate (Figure \ref{arr_rates}) derived from the Alibaba data center traces\cite{weng2022mlaas}. Since the traces do not reveal whether a job is an inference or training machine learning job, we classify each job as either inference or training according to a uniform distribution, and we base duration assumptions on this classification.  The duration of inference jobs is exponentially distributed with a lambda value of 3, and the duration of training jobs is uniformly distribution with values between 10 and 40. Each job is categorized into one of three equally likely performance types: linear, sublinear, or capped, illustrated in Figure \ref{throughpur_char}. 
As defined before, a linear performance type means that a job’s duration on different slices scales linearly. A sublinear performance type means that a job’s duration scales sub-linearly on different slices. There are four different sublinear functions simulated as exponential and logarithmic functions, each with an equal likelihood of occurrence. A capped performance type means that a job’s duration scales linearly to a certain slice, at which point its performance does not improve despite increased resources of larger slices. A job’s performance can be capped at 2g, 3g, or 4g slices, modelled to be selected uniformly at random. Lastly, the LLF and LALF algorithms have critical laxity thresholds and the maximum number of critical laxity preemptions set to 1 and 3 respectively. 

Experiments on the repartitioning framework consist of queues simulated over 24 hours and repeated for 500 iterations. We benchmark against three other types of solutions. The first benchmark is a full A100 40GB GPU without MIG enabled, where it is important to note the difference in performance between using a full MIG-enabled GPU with a 7g.40gb slice and using the MIG-disabled GPU we use for our benchmark: using the full GPU without MIG, there is a 6\% performance improvement for linear jobs using 250W compared to a 7g.40gb slice on a MIG-enabled GPU [22]. The second benchmark is an A100 40GB GPU with MIG enabled on a fixed configuration without re-partitioning. We tested every possible fixed configuration and determined that configuration 3 presented the strongest \metric~ results as the selected fixed configuration. The third and final benchmark is an A100 40GB GPU with MIG enabled on two fixed configurations similarly set to Configuration 6 for day-time, defined as 5:00-17:00, and Configuration 2 for night-time, from 17:00-5:00, due to varying arrival rates between these time intervals. 


\subsection{In-Configuration Scheduling Algorithm Analysis}

EDF-SS was the strongest scheduling algorithm out of the four evaluated according to \metric~. Relative to the other algorithms, EDF-SS  generally has lower energy consumption with an average tardiness similar to the others. We analyze each algorithm's behavior across different arrival rates, inference-training splits, and configurations to determine the circumstances in which EDF-SS outperforms the others and when it might not. 

We see EDF-SS either match or outperform the other algorithms on configurations 1, 2, 5, 8, 9, 10, and 11. Some of these configurations such as configurations 1 and 5 are made up of one sized slice (7g.40gb for the former and two 3g.20gb slices for the latter), where different slice allocation policies result in the same scheduling decision between algorithms and only job prioritization decisions differentiate them; EDF variants produce the same value, as do the LLF variants, but LLF will prioritize longer-running jobs with lesser laxity and push a greater number of shorter jobs past their deadlines. For configurations 8, 9, 10, and 11 that have many smaller sized slices, EDF-SS more strategically uses the 1g.10gb and 1g.5gb slices by intentionally selecting and allocating jobs to them more often than the other algorithms; there is a higher rate for EDF-SS of 1g.10gb and 1g.05gb slices being selected while also being the natural first-choice slice for jobs. The other configurations where EDF-SS does not have the minimum performance metric, including configurations 3, 4, 6, and 7, have larger and fewer slices. EDF-FS performed the best in these configurations because it consistently had the lowest average tardiness. For these configurations the LLF variants would preempt more frequently at critical laxity points, so although their job prioritization decisions differed from EDF variants, they would re-prioritize jobs more frequently to meet their deadlines. This is unlike the EDF algorithms that have less flexibility to preempt,  so we see that EDF-SS and EDF-FS are often the worst and best-performing algorithms in these circumstances with the LLF variants managing to stay in between for this reason. 

\begin{figure}
\centering
  \includegraphics[width=\linewidth]{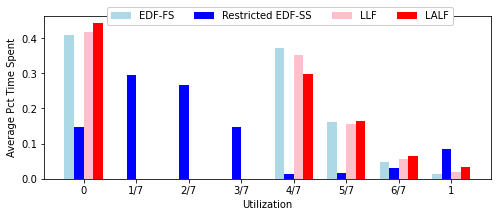}
  \caption{\metric~ Average Percentage Time Spent Per Utilization Level}
  \label{fig:util}
\end{figure}

\begin{figure}
\centering
  \includegraphics[width=\linewidth]{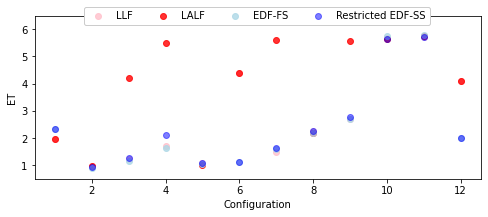}
  \caption{\metric~ Across Configurations: Arrival Rate 0.1}
  \label{fig:a01}
\end{figure}

\begin{figure}
\centering
  \includegraphics[width=\linewidth]{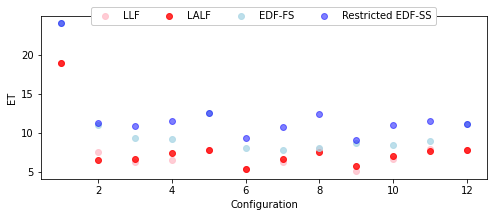}
  \caption{\metric~ Across Configurations: Arrival Rate 0.75}
  \label{fig:a075}
\end{figure}

When we reduce the arrival rate from $0.5$ to $0.1$ as shown in Figure \ref{fig:a01}, EDF-SS generally improves compared to other algorithms. Analyzing the average utilization of the algorithms per configuration across their simulations in \ref{fig:util} 
reveals that, as an example on configuration 4 composed of one 4g20gb slice and three 1g10gb slices, the other algorithms rarely spend significant time on lower utilization levels below $4/7$ of the GPU; they prioritize scheduling the larger slices such as 4g20gb first, while EDF-SS will slow down processing times and utilize the smaller slices simultaneously and thus spends significant time in lower utilization. However, this behavior can negatively impact EDF-SS with its energy efficiency for mid-range arrival rates, and especially with its tardiness for much greater arrival rates.  We experimented with a very aggressive arrival rate of $0.75$, which is much higher than the greatest arrival rate observed in Alibaba's dataset, to understand how the algorithms behave in this setting. The results depicted in Figure \ref{fig:a075} show that while some jobs were either very difficult to schedule or even impossible to schedule, none of the algorithms perform particularly well, and EDF-SS performs the worst. With a heavier load, longer-running higher-priority jobs that were initially allocated to slower slices and expected to complete their deadline would often be re-prioritized and no longer be able to meet their deadlines on time, leading to an increase in both  maximum tardiness and average tardiness.




\begin{figure}
\centering
  \includegraphics[width=\linewidth]{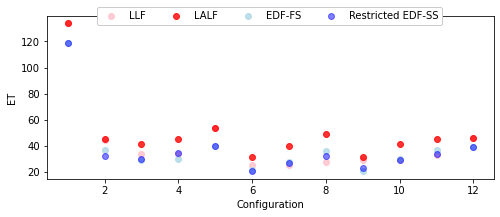}
  \caption{\metric~ Across Configurations: Inference Split 20\%}
  \label{fig:inf20c}
\end{figure}

\begin{figure}
\centering
  \includegraphics[width=\linewidth]{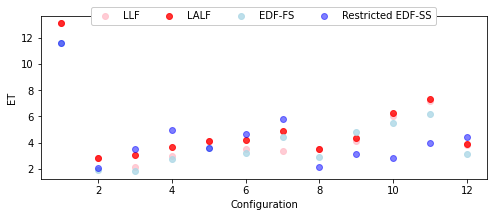}
  \caption{\metric~ Across Configurations: Inference Split 80\%}
  \label{fig:inf80c}
\end{figure}

When we vary the inference and training split, EDF-SS again further improves compared to other algorithms. Evaluating the results of a $20\%$ inference split as seen in Figure \ref{fig:inf20c}, EDF-SS either outperforms or almost ties EDF-FS according to \metric~. With a greater number of longer-running jobs and in the way an increase in load, all algorithms struggled with average tardiness compared to the original $80\%$ inference split as shown in Figure \ref{fig:inf80c}, but EDF-SS schedules were most energy efficient; The system would often have at least one job running, EDF-SS had the greatest amount of time spent processing at least one job, while the other algorithms would more quickly finish the longer-running training jobs and reduce the opportunity for overlap that EDF-SS capitalized off of. Additionally, this slowest slice allocation paired specifically with the earliest deadline job prioritization, as opposed to least laxity, ensures timely completion of shorter jobs; least laxity can prioritize longer-running jobs with lower laxities, potentially creating a cascading effect where subsequent jobs miss their deadlines, leading to higher average tardiness in such cases.

EDF-SS was the strongest overall algorithm candidate for in-configuration scheduling, but it has weaknesses during peak arrival rates and on configurations with fewer and faster slices. In the next section, we will understand how the RL agent might choose certain configurations that not only fit the circumstances of the queue, but also the known strengths and weaknesses of EDF-SS. 


\subsection{Dynamic Repartitioning Analysis}

We evaluate the results of the dynamic repartitioning model against three benchmarks: No MIG, Static MIG, and Day/Night Repartitioning MIG. These benchmarks are chosen to provide realistic comparisons to existing scheduling approaches with MIG; while MIG currently can technically be repartitioned, implementing dynamic repartitioning is challenging due to the lack of an orchestration framework to facilitate the process. The industry-standard container orchestration platform, Kubernetes, has recently proposed the development of such a framework; however, until then, currently implemented MIG scheduling approaches cannot easily assume frequent or dynamic repartitioning. To address this, we include a twice-daily repartitioning benchmark to provide a more rigorous yet realistic evaluation. The benchmarks also aim to highlight the value of MIG and demonstrate how dynamically adjusting partitions is essential for fully realizing its benefits. Our findings indicate a significant improvement in the dynamic repartitioning model over each benchmark based on \metric, and our analysis will explain how varying repartitioning rates throughout the day and preferred configurations contribute to this improvement.

\begin{table}[ht]
  \caption{Results using Dynamic Repartitioning}
  \label{multi_config}
  \begin{center}
    \begin{tabular}{p{3.5cm}|p{3.5cm}}
        \hline
        Algorithm & \metric~ \\
        \hline
        No MIG & 8.56 \\
        Static MIG & 3.97 \\
        Day-Night MIG &  3.68 \\
        Dynamic MIG & 2.82 \\
        \hline
    \end{tabular}
  \end{center}
\end{table}

Dynamic MIG significantly outperforms each benchmark model as shown in Table \ref{multi_config}. Its advantages over No MIG are particularly noticeable when running jobs that cannot fully utilize the GPU; for instance, capped jobs will not process faster beyond a certain resource threshold even if it is allocated more. In such periods, Dynamic MIG can effectively pack the GPU, leading to improved energy efficiency and tardiness compared to each benchmark. During peak periods with sublinear and capped jobs, No MIG can inefficiently consume up to twice the energy and cause avoidable tardiness for jobs that could have been executed concurrently on smaller slices in less time. 

Additionally, Dynamic MIG outperforms Static MIG, especially in scenarios with multiple linear jobs or sublinear and capped jobs that scale to larger slices, such as 4g.20gb.
In these cases, alternative configurations with larger slices, like configurations 1 or 2, can complete jobs on time, while configuration 3 may lead to tardiness, particularly under tight deadlines. In these cases, it is evident that specific configurations are better suited to different mixes of jobs with both linear and nonlinear performance scales. 

Analyzing the repartitioning frequency reveals that Dynamic MIG particularly excels during the peak hours of 5:00 to 17:00, corresponding to a non-homogeneous arrival rate in Figure \ref{arr_rates}. Repartitioning also occurs consistently during transition idle periods of 3:00 to 5:00 and 17:00 to 19:00 as the system prepares for anticipated loads. 

\begin{figure}
    \centering
    \includegraphics[width=\linewidth]{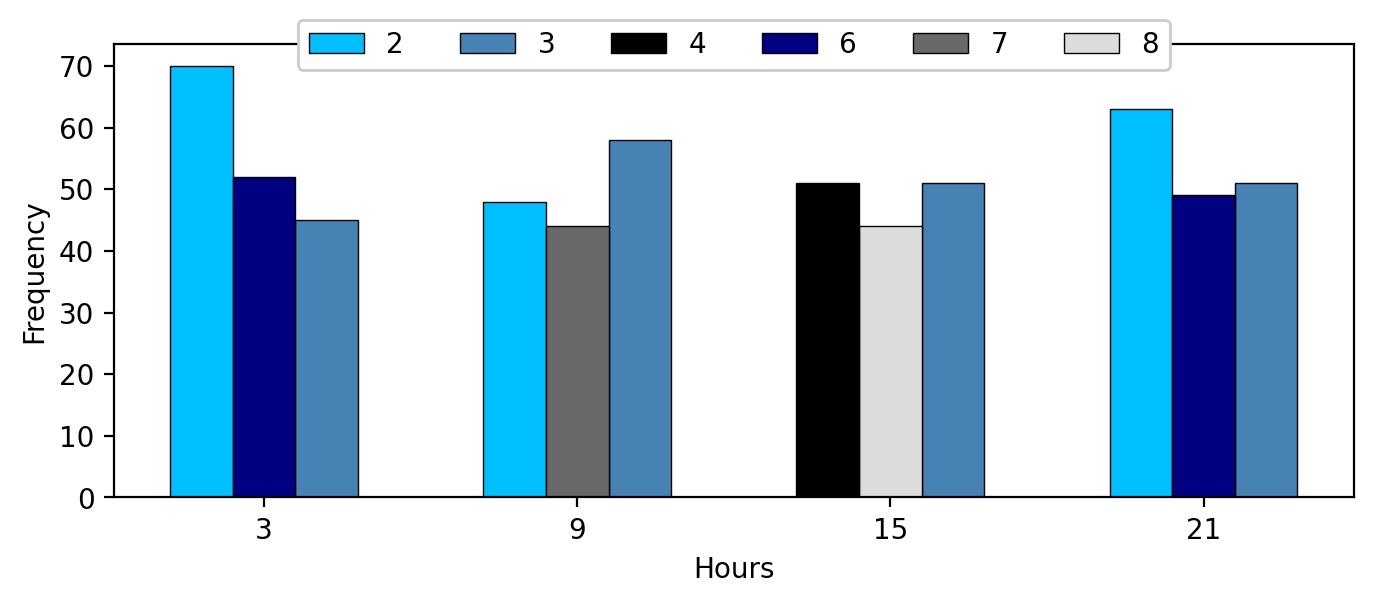}
     \mbox{ }
    \caption{Preferred Configurations by 4-Hour Interval}
    \label{pop_configur}
\end{figure}

The RL model favors different configurations under various circumstances, including at different times of day as shown in Figure \ref{pop_configur}. Overall, configurations 2, 3, and 6 are the most commonly used throughout the day. During the 0:00-6:00 and 18:00-23:59 intervals, configuration 2 and its two larger 4g.20gb and 3g.20gb slices are most consistently selected . Configurations with multiple 1g.10gb or 1g.5gb slices are less commonly seen and only occur if that same configuration has larger slices as well, such as in configuration 3 which also features a 4g20gb slice. The preferred configurations during peak hours (configurations 2, 3, 4, 7, and 8) are diverse, with configurations 2 and 3 having fewer and typically faster slices, while configurations 8 and 10 have a greater number of smaller slices, and configuration 4 sits in between the two. Given the relatively low cost of repartitioning, the model tends to adopt configurations that align better with current job demands rather than sticking with a less suitable configuration. Remaining on the same configuration typically occurs during overnight and early morning hours when the selected configuration already optimally serves the typically few jobs in the queue. 

In conclusion, our analysis shows that Dynamic MIG outperforms a full GPU setup and other MIG approaches by improving both performance and energy efficiency. By adapting to varying workloads and effectively selecting configurations based on job requirements, Dynamic MIG more efficiently uses the GPU and minimizes tardiness relative to the benchmark models. This flexibility is crucial for meeting the diverse demands of jobs within a data center environment.



\section{Conclusion and Future Work}

We leveraged MIG capabilities, the GPU's power profile, and inelastic workloads to jointly minimize energy consumption and improve performance. First, we considered static MIG configurations and contrasted four scheduling disciplines with various slice allocation and job prioritization approaches. Overall, the EDF-SS scheduling discipline performed better using a combined measure of energy and tardiness. Next, we considered dynamically repartitioning the GPU to adapt to workload fluctuates throughout the day by employing reinforcement learning to select the configurations. Realistic workload arrival patterns revealed that specific configurations are preferred at different times, supporting a policy for repartitioning and predictive slicing.

Dynamic Repartitioning improved both the energy efficiency and average tardiness of scheduling inference and fine-tuning simulated workloads with a mix of performance types when compared to an otherwise Static MIG, No MIG, or Day-Night MIG solutions. Specifically Dynamic Repartitioning provided a $26\%$, $31\%$ and $68\%$ improvement in \metric~ over each benchmark respectively. By enabling repartitioning with MIG, it is possible to better utilize the GPU and to take advantage of the energy opportunity of an A100 MIG with mixed-performance job types. 

Further work on this topic might improve the heuristic dependency design and explore partial repartitioning of the GPU or nonstandard configuration choices. 
could also offer further improvements to the \metric~ in our model. This work considers the scheduling of training and inference jobs on a single GPU.
 Our ultimate goal is to understand how to schedule a whole data center, which consists of multiple machines of different types.  Our work, which schedules across multiple slices on one GPU, considers this limited form of parallelism with now a foundational understanding of energy-awareness on MIG, and we plan to consider the much greater parallelism opportunities afforded by scheduling on a whole data center.

 \section*{Acknowledgment}
This research is partially supported by funding from IBM. We would like to thank Pedro Bello-Maldonado for his help with the early stages of this research.

\bibliographystyle{ieeetr}
\bibliography{cite}

\end{document}